\documentclass[useAMS,usenatbib]{mn2e}
\usepackage{graphicx}
\usepackage{epstopdf}
\usepackage{amsmath}
\usepackage{amssymb}
\title[Mass accretion rate]{What is the real accretion rate onto a black hole for low angular momentum accretion?}
\author[Bu \& Yang]{De-Fu Bu$^1$\thanks{E-mail: dfbu@shao.ac.cn (DB)} and Xiao-Hong Yang$^2$\thanks{E-mail: yangxh@cqu.edu.cn} \\
$^{1}$Key Laboratory for Research in Galaxies and Cosmology, Shanghai Astronomical Observatory, Chinese Academy of Sciences,\\ 80 Nandan Road, Shanghai 200030, China \\
$^2$ Department of Physics, Chongqing University, Chongqing 400044, China}
\begin{document}

\pagerange{\pageref{firstpage}--\pageref{lastpage}} \pubyear{2002}

\maketitle

\label{firstpage}

\begin{abstract}
Mass accretion rate is a key parameter in accretion disk theory. It determines black hole accretion mode. In large scale cosmological simulations studying galaxy formation and evolution, Bondi radius can at most be marginally resolved. In those simulations, Bondi accretion formula is always used to estimate black hole accretion rate. Bondi solution is too simple, which cannot represent the real accretion process. We perform simulations of hot accretion flow at parsec scale irradiated by a low luminosity active galactic nucleus (AGN). We perform 77 simulations with varying density and temperature at outer boundary (10 parsec). Our purpose is to find a formula to calculate real black hole accretion rate based on the gas density and temperature at parsec scale. We define Eddington accretion rate to be $\dot M_{\rm Edd}=10L_{\rm Edd}/c^2$, with $L_{\rm Edd}$ and $c$ been Eddington luminosity and speed of light respectively. We set black hole mass to be $10^8M_{\odot}$, $M_{\odot}$ is solar mass. We find that black hole accretion rate can be expressed as $\frac{\dot M_{\rm BH} }{\dot M_{\rm Edd}}=10^{-3.11} (\rho_0/10^{-22}{\rm g cm}^{-3})^{1.36} (T_0/10^7 {\rm K})^{-1.9}$, with $\rho_0$ and $T_0$ being density and temperature at parsec scale, respectively. We find the formula can accurately predict the luminosity of observed low-luminosity AGNs (with black hole mass $\sim 10^8M_{\odot}$). This formula can be used in the sub-grid models in large scale cosmological simulations with a black hole mass of $\sim 10^8M_{\odot}$.
\end{abstract}

\begin{keywords}
accretion, accretion disks -- black hole physics -- hydrodynamics -- galaxies: active -- galaxies: nuclei.
\end{keywords}

\section{Introduction}
Mass accretion rate is a key parameter in black hole accretion physics.
It determines the accretion mode of the black hole. When the accretion rate
is lower than $2\%$ of $\dot M_{\rm Edd}$, the black hole accretes gas by hot accretion
flow (Yuan \& Narayan 2014). When the accretion rate is in the range
$2\%\dot M_{\rm Edd} < \dot M < \dot M_{\rm Edd}$, black hole accretes gas by standard
thin disk (Shakura \& Sunyaev 1973). When the accretion rate is higher than the Eddington rate,
black hole accretes gas by slim disk (Abramowicz et al. 1988). Different accretion modes have
quite different spectrum.

Almost every galaxy hosts a super massive black hole at its center.
The tight correlation between black hole mass and various properties of its host galaxy
indicates that the central black hole and its host
galaxy co-evolve (e.g., Magorrian et al. 1998; Ferrarese \& Merritt 2000; Gebhardt et al. 2000;
Tremaine et al 2002; H\"{a}ring \& Rix 2004; G\"{u}ltekin et al. 2009; Kormendy \& Ho 2013).
AGN feedback plays an important role in the evolution of its host galaxy (Fabian 2012).

Both radiation and wind from an AGN can interact with its host galaxy. The AGN radiation can heat the interstellar
medium (ISM; Ciotti \& Ostriker 1997; 2001; 2007; Ciotti et al. 2009). The heating efficiency depends on the luminosity and Compton temperature of the photons emitted by the AGN. The luminosity of accretion system depends on the black hole accretion rate. The Compton temperature of photons depends on the accretion mode of black hole, which depends on the accretion rate again. For low-luminosity active galactic nucleus (LLAGN), hot accretion flow operates.
Most of the photons emitted by LLAGNs are in X-ray band, and the Compton temperature of photons is $10^8 {\rm K}$ (Xie et al. 2017). For a quasar, cold standard thin disk operates. Most of the emitted photons by quasar are in ultraviolet (UV) band, and the Compton temperature of photons is $2 \times 10^7 {\rm K}$ (Sazonov et al. 2004). The radiative heating efficiencies by LLAGNs and quasars are quite different due to different luminosities and Compton temperatures.

In addition to radiative heating, radiation pressure can also affect the properties of ISM. For a LLAGN, the photons are mainly in X-ray band. The radiation pressure is just due to Compton scattering. Because the luminosity of a LLAGN is much lower than the Eddington luminosity, the radiation pressure is much smaller than the gravity. However, for a quasar, the line force due to interaction between UV photons and the not fully ionized gas can exceed gravity significantly. Strong wind can be driven by line force (e.g., Proga et al. 2000; Murray et al. 1995; Murray \& Chiang 1997; Kurosawa \& Proga 2009; Liu et al. 2013; Nomura et al. 2016; Nomura \& Ohsuga 2017).

Wind (or outflow) from an AGN can also interact with ISM effectively (e.g., Ostriker et al. 2010; Wang et al. 2010; Weinberger et al. 2017; Yuan et al. 2018). Blue shifted absorption lines are frequently detected in luminous AGNs (e.g., Crenshaw et al. 2003; Tombesi et al. 2010, 2014; King \& Pounds 2015; Liu et al. 2015; Gofford et al. 2015) and black hole X-ray binaries (BXBs) (e.g., Neilsen \& Homan 2012; Homan et al. 2016; D\'{i}az Trigo \& Boirin 2016). These observations indicate that winds are common phenomenon in standard thin disk powered Luminous AGNs and BXBs. It is very hard to directly observe winds in LLAGNs and hard state of BXBs. The reason may be as follows. LLAGNs and the hard state of BXBs are powered by hot accretion flow. Hot accretion flow is fully ionized. Therefore, it is hard to detect absorption lines. Fortunately, in recent years, we have gradually accumulated indirect evidences that wind can also be generated in hot accretion flow (e.g., Crenshaw \& Kramemer 2012; Wang et al. 2013; Cheung
et al. 2016; Homan et al. 2016). The properties of wind from hot accretion flow are mainly investigated by numerical simulations (e.g., Tchekhovskoy et al. 2011; Yuan et al. 2012, 2015; Narayan et al. 2012; Li et al. 2013; see also Moller \& Sadowski 2015) and analytical works (e.g., Cao 2011; Wu et al. 2013; Gu 2015). The properties of winds from luminous AGNs and LLAGNs are quite different. For example, the velocity of wind in luminous AGNs is proportional to luminosity (or equivalently mass accretion rate) of accretion disk (Gofford et al. 2015). However, the velocity of wind from LLAGNs is independent of luminosity of accretion flow. The velocity of wind from LLAGNs is only determined by the location where winds are generated (Yuan et al. 2015). The temperature (or thermal power) of winds from LLAGNs and luminous AGNs is also quite different. Due to the different properties of wind from LLAGNs and luminous AGNs, we can expect that wind feedback at different accretion mode can be quite different.

From the introduction above, we know different accretion mode can have very different feedback. The accretion mode is determined by the black hole accretion rate. However, in large scale cosmological simulations studying galaxy formation and evolution, the Bondi radius can at most be marginally resolved (e.g., Springel et al. 2005; Booth \& Schaye 2009; McCarthy et al. 2010, 2017; Teyssier et al. 2011; Choi et al. 2012; 2015; Gan et al. 2014; Le Brun et al. 2014; Vogelsberger et al. 2014; Khandai et al. 2015; Schaye et al 2015; Negri \& Volonteri 2017). In these simulations, the authors use the Bondi solution to estimate the black hole accretion rate. The black hole growth and accretion mode are determined by the Bondi solution. Recently, Bondi solutions including galaxy gravitational potential have been developed (Korol et al. 2016; Ciotti \& Pellegrini 2017; Ciotti \& Pellegrini 2018). The Bondi accretion model is simple but has problems. First, the Bondi solution assumes spherical symmetry. However, the real accretion flow can not be spherical symmetry because gas has angular momentum. Second, the Bondi solution assumes that the accretion rate does not change from the Bondi radius down to the central black hole. In reality, due to the presence of wind, the accretion rate decreases from the Bondi radius towards the black hole. Some previous works (e.g., Igumenshchev \& Narayan 2002; Moscibrodzka 2006; Gaspari et al. 2013; Gaspari et al. 2017) have shown that Bondi
formula is not good estimate to calculate black hole accretion rate. However, no papers have given
a formula to calculate the real black hole accretion rate based on gas density and temperature at
Bondi radius. In this paper, we are the first to give a formula to calculate black hole accretion rate based on density and temperature at Bondi radius.

Yang \& Bu (2018) studied the slowly rotating accretion flow in the region from hundreds of $r_s$ ($r_s$ is Schwarzschild radius) to the region beyond the Bondi radius. The accretion flow studied by Yang \& Bu (2018) connects the accretion flow very close to the black hole to the flow on galactic scale. Yang \& Bu (2018) studied the low accretion rate flow which is irradiated by a LLAGNs. We find that the real accretion rate onto the black hole deviates from the Bondi value significantly. In this paper, we study the slowly rotating low accretion rate flow irradiated by a LLAGN in the region from hundreds of $r_s$ to $10$ parsec (hereafter, pc for short). This paper differs from that of Yang \& Bu (2018) in two aspects. First, in Yang \& Bu (2018), only the black hole gravity is taken into account. However, in the region around 10 parsec, the gravitational force of nuclear stars cluster is comparable to that of the central black hole. The stars gravity may be important and needs to be taken into account. In the present paper, we will take into account the gravity of nuclear stars cluster. Second, in Yang \& Bu (2018), we find with the changing of density and temperature at parsec scale, the central black hole accretion rate can be very different. However, the number of runs of simulations in Yang \& Bu (2018) is very small. In this paper, we set the black hole mass to be $M=10^8M_{\odot}$. We run 77 simulations covering a large parameter space of densities and temperatures at the outer boundary. Our purpose is to find a analytical formula to calculate the real accretion rate onto the black hole as a function of density and temperature at parsec scale. We do find a analytical formula. We find that the analytical formula can accurately predict the luminosity of observed LLAGNs with a $\sim 10^8M_{\odot}$ black hole mass. This is very useful to the large scale simulations studying galaxy formation and evolution.

In section 2, we describe our models and method; In section 3, we present our results; Section 4
is devoted to summary and discussion.

\section{Numerical method }

We employ spherical coordinates ($r,\theta,\phi$) and use ZEUS-MP code (Hayes et al. 2006) to solve the equations below:

\begin{equation}
 \frac{d\rho}{dt} + \rho \nabla \cdot {\bf v} = 0,
\end{equation}
\begin{equation}
 \rho \frac{d{\bf v}}{dt} = -\nabla p - \rho \nabla \Phi 
\end{equation}
\begin{equation}
 \rho \frac{d(e/\rho)}{dt} = -p\nabla \cdot {\bf v} + \dot E
\end{equation}
$\rho$, $\bf v$ and $p$ are density, velocity and internal energy, respectively. For gas pressure, we have $p=(\gamma-1)e$. We set $\gamma=5/3$.

As introduced above, we take into account both the gravity potential of the central black hole and the nuclear stars cluster. For the black hole potential, we have $\Phi_{\rm BH}=-GM/(r-r_s)$ where $M$ and $G$ are the central black hole mass and the gravitational constant, respectively. The Schwarzschild radius $r_s=2GM/c^2$. In this paper, we set the black hole mass to be $M=10^8M_{\odot}$. The velocity dispersion of nuclear stars is assumed to be a constant for varying radius. This seems to be the case for many AGNs (e.g., Kormendy \& Ho 2013). Therefore, the gravitational potential of the nuclear stars is $\Phi_{\rm star}= \sigma^2 \ln (r) + C$. $\sigma$ is the velocity dispersion of stars and $C$ is a constant. Kormendy \& Ho (2013) find that the typical value of $\sigma$ is $(100-400)$ km s$^{-1}$. It has been shown that for a $10^8M_{\odot}$ black hole, the stellar velocity dispersion is in the range $150-250$ km s$^{-1}$ (Greene \& Ho 2006). In this paper, we set $\sigma=200$ km s$^{-1}$. In radial direction, our computational domain is $500 r_{\text{s}} \leq r \leq 10^6 r_{\text{s}}$ ($5\times 10^{-3}$ pc $\leq r \leq 10$ pc). At the outer boundary (10pc), the stars gravity is 0.9 times the gravity of the black hole. From the outer boundary to smaller radii, the black hole gravity increases faster than that of the stars. Therefore, in the whole computational domain, the black hole gravity dominates. At the outer boundary, the stars gravity is comparable to that of the black hole. Because our outer boundary is 10pc, therefore, we neglect the gravity of dark matter.

The upper limit of luminosity of a LLAGN is 2$\% L_{\rm Edd}$ (e.g., Yuan \& li 2011). In this paper, we consider LLAGNs with luminosity lower than 2$\% L_{\rm Edd}$. In the energy equation, $\dot{E}$ is gas net heating and cooling rate. We take into account Compton heating/cooling, bremsstrahlung cooling, photoionization heating, and line and
recombination cooling. Xie et al. (2017) find that for a LLAGN, the Compton temperature $T_X$ of the photons is $\sim 10^8$K. In this paper, we set $T_X=10^8$K. For the detailed formula of heating and cooling, we refer to Sazonov et al. (2005).

Initially, the whole computational domain is filled with gas with uniform density ($\rho_0$) and temperature ($T_0$). The angular momentum of gas is equal to the Keplerian angular momentum at $350r_{\text{s}}$.  In order to achieve a steady state, we also inject low angular momentum gas into the computational domain. The angular momentum of injected gas is equal to the Keplerian angular momentum at $350r_{\text{s}}$. The injected gas has density $\rho=\rho_0$ and temperature $T=T_0$. The computational domain and properties of injected gas in this paper are same as those in Bu \& Yang (2018) and Yang \& Bu (2018). We refer to Yang \& Bu (2018) for the details of calculation of black hole accretion rate and luminosity.

Our computational domain is $0 \leq \theta \leq \pi/2$ with 88 grids uniformly spaced in $\theta$ direction. We have $500 r_{\text{s}} \leq r \leq 10^6 r_{\text{s}}$ and 140 grids logarithmcally spaced ($dr_{i+1}/dr_i=1.05$) in $r$ direction. Axis-of-symmetry and reflecting boundary conditions are applied at
the pole and the equatorial plane, respectively. Outflow boundary condition is
adopted at the inner radial boundary. At outer radial boundary, if gas is inflowing, we set density $\rho=\rho_0$ and temperature $T=T_0$. If gas is outflowing, we use outflow boundary conditions.

\begin{table*} \caption{Simulation parameters and results }
\setlength{\tabcolsep}{4mm}{
\begin{tabular}{ccccccc}
\hline \hline
 Model & Stars gravity & $\rho_0$                    & $T_0$   & Bondi radius ($R_B$)   &  $\dot M_{\rm in}(r_{\rm in})$ & $\dot M_{\rm BH}$\\

  &   & ($10^{-22}\text{g cm}^{-3}$)& ($10^7$K) & ($10 {\rm pc}$)  & ($L_{\rm Edd}/0.1c^2$)         & ($L_{\rm Edd}/0.1c^2$)   \\
(1) & (2)             & (3)                         &  (4)  & (5)  & (6)    &        (7)      \\

\hline\noalign{\smallskip}
SD263T0.1 & ON & $3\times 10^{-4}$ & 0.1 & $1.6$ & $3.15\times10^{-6}$ & $5.32\times 10^{-7}$ \\
SD263T0.2 & ON & $3\times 10^{-4}$ & 0.2 & $0.8$ & $2.1 \times10^{-6}$ & $3.55\times 10^{-7}$ \\
SD263T0.4 & ON & $3\times 10^{-4}$ & 0.4 & $0.4$ & $1.04\times10^{-6}$ & $1.75\times 10^{-7}$ \\
SD263T0.6 & ON & $3\times 10^{-4}$ & 0.6 & $0.27$& $4.13\times10^{-7}$ & $6.98\times 10^{-8}$ \\
SD263T0.8 & ON & $3\times 10^{-4}$ & 0.8 & $0.2$ & $2.69\times10^{-7}$ & $4.55\times 10^{-8}$ \\
SD263T1   & ON & $3\times 10^{-4}$ &   1 & $0.16$& $9.64\times10^{-8}$ & $1.63\times 10^{-8}$ \\
SD263T2   & ON & $3\times 10^{-4}$ &   2 & $0.08$& $3.56\times10^{-8}$ & $6.02\times 10^{-9}$ \\
SD263T4   & ON & $3\times 10^{-4}$ &   4 & $0.04$& $7.57\times10^{-9}$ & $1.28\times 10^{-9}$ \\
SD263T6   & ON & $3\times 10^{-4}$ &   6 & $0.027$&  $4.85\times10^{-9}$ & $8.2 \times 10^{-10}$ \\
SD263T0.1 & ON & $3\times 10^{-4}$ &   8 & $0.02$ & $2.78\times10^{-9}$ & $4.7 \times 10^{-10}$ \\
SD25T0.1 & ON & $1\times 10^{-3}$  & 0.1 & $1.6$& $1.26\times10^{-5}$ & $2.14\times 10^{-6}$ \\
SD25T0.1 & ON & $1\times 10^{-3}$  & 0.2 & $0.8$ & $6.5 \times10^{-6}$ & $1.1\times 10^{-6}$ \\
SD25T0.1 & ON & $1\times 10^{-3}$  & 0.4 & $0.4$ & $2.98\times10^{-6}$ & $5.03\times 10^{-7}$ \\
SD25T0.1 & ON & $1\times 10^{-3}$  & 0.6 & $0.27$ & $1.2 \times10^{-6}$ & $2.03\times 10^{-7}$ \\
SD25T0.1 & ON & $1\times 10^{-3}$  & 0.8 & $0.2$ & $4.61\times10^{-7}$ & $7.8 \times 10^{-8}$ \\
SD25T0.1 & ON & $1\times 10^{-3}$  & 1  & $0.16$  & $3.49 \times10^{-7}$ & $5.9 \times 10^{-8}$ \\
SD25T0.1 & ON & $1\times 10^{-3}$  & 2  & $0.08$   & $9.05 \times10^{-8}$ & $1.53\times 10^{-8}$ \\
SD25T0.1 & ON & $1\times 10^{-3}$ & 4   & $0.04$   & $4.07 \times10^{-8}$ & $6.88 \times 10^{-9}$ \\
SD25T0.1 & ON & $1\times 10^{-3}$ & 6   & $0.027$   & $1.41 \times10^{-8}$ & $2.38\times 10^{-9}$ \\
SD25T0.1 & ON & $1\times 10^{-3}$ & 8 & $0.02$   & $1.05 \times10^{-8}$ & $1.78\times 10^{-9}$ \\
SD253T0.1 & ON & $3\times 10^{-3}$ & 0.1  & $1.6$ & $4.22\times10^{-5}$ & $7.14\times 10^{-6}$ \\
SD253T0.2 & ON & $3\times 10^{-3}$ & 0.2 & $0.8$ & $1.53\times10^{-5}$ & $2.59\times 10^{-6}$ \\
SD253T0.4 & ON & $3\times 10^{-3}$ & 0.4 & $0.4$ & $7.45\times10^{-6}$ & $1.26\times 10^{-6}$ \\
SD253T0.6 & ON & $3\times 10^{-3}$ & 0.6 & $0.27$ & $3.66\times10^{-6}$ & $6.2\times 10^{-7}$ \\
SD253T0.8 & ON & $3\times 10^{-3}$ & 0.8 & $0.2$ & $2\times10^{-6}$ & $3.39\times 10^{-7}$ \\
SD253T1 & ON & $3\times 10^{-3}$ & 1 & $0.16$ & $9.94\times10^{-7}$ & $1.68\times 10^{-7}$ \\
SD253T2 & ON & $3\times 10^{-3}$ & 2 & $0.08$ & $5.63\times10^{-7}$ & $9.52\times 10^{-8}$ \\
SD253T4 & ON & $3\times 10^{-3}$ & 4 & $0.04$ & $1.12\times10^{-7}$ & $1.9\times 10^{-8}$ \\
SD253T6 & ON & $3\times 10^{-3}$ & 6 & $0.027$ & $4.02\times10^{-8}$ & $6.8\times 10^{-9}$ \\
SD253T8 & ON & $3\times 10^{-3}$ & 8 & $0.02$ & $2.45\times10^{-8}$ & $4.14\times 10^{-9}$ \\
SD24T0.1 & ON & 0.01      &   0.1 & $1.6$   & $3.4 \times10^{-4}$ & $5.7  \times 10^{-5}$ \\
SD24T0.2 & ON & 0.01      &   0.2 & $0.8$   & $6.4 \times10^{-5}$ & $1.08 \times 10^{-5}$ \\
SD24T0.4 & ON & 0.01      &   0.4 & $0.4$   & $2.4 \times10^{-5}$ & $4.1  \times 10^{-6}$ \\
SD24T0.6 & ON & 0.01      &   0.6 & $0.27$   & $1.47\times10^{-5}$ & $2.48 \times 10^{-6}$ \\
SD24T0.8 & ON & 0.01      &   0.8 & $0.2$   & $4.63\times10^{-6}$ & $7.8  \times 10^{-7}$ \\
SD24T1   & ON & 0.01      &   1 & $0.16$     & $3.6 \times10^{-6}$ & $6.1  \times 10^{-7}$ \\
SD24T2   & ON & 0.01      &   2 & $0.08$     & $2.1 \times10^{-6}$ & $3.5  \times 10^{-7}$ \\
SD24T4   & ON & 0.01      &   4 & $0.04$     & $3   \times10^{-7}$ & $5.1  \times 10^{-8}$ \\
SD24T6   & ON & 0.01      &   6 & $0.027$     & $1.88\times10^{-7}$ & $3.2  \times 10^{-8}$ \\
SD24T8   & ON & 0.01      &   8 & $0.02$     & $8.6 \times10^{-8}$ & $1.45 \times 10^{-8}$ \\
SD243T0.1 & ON & 0.03         &   0.1 & $1.6$     & $2\times10^{-3}$ & $3.4  \times 10^{-4}$ \\
SD243T0.2 & ON & 0.03         &   0.2 & $0.8$     & $1\times10^{-3}$ & $1.7  \times 10^{-4}$ \\
SD243T0.4 & ON & 0.03         &   0.4 & $0.4$     & $6\times10^{-5}$ & $1.01  \times 10^{-5}$ \\
SD243T0.6 & ON & 0.03         &   0.6 & $0.27$     &$5.8\times10^{-5}$ & $9.8  \times 10^{-6}$ \\
SD243T0.8 & ON & 0.03         &   0.8 & $0.2$     &$3.35\times10^{-5}$ & $5.67  \times 10^{-6}$\\
SD243T1 & ON &   0.03         &   1 & $0.16$       &$1.58\times10^{-5}$ & $2.67  \times 10^{-6}$ \\
SD243T2 & ON &   0.03         &   2 & $0.08$       &$4.6 \times10^{-6}$ & $7.8   \times 10^{-7}$ \\
SD243T4 & ON &   0.03         &   4 & $0.04$       &$7.4 \times10^{-7}$ & $1.25  \times 10^{-7}$ \\
SD243T6 & ON &   0.03         &   6 & $0.027$       &$4.5 \times10^{-7}$ & $7.6   \times 10^{-8}$ \\
SD243T8 & ON &   0.03         &   8 & $0.02$       &$3   \times10^{-7}$ & $5.1   \times 10^{-8}$ \\
SD23T0.2 & ON & 0.1       &   0.2 & $0.8$   & $6.75\times10^{-3}$ & $1.14 \times 10^{-3}$\\
SD23T0.4 & ON & 0.1       &   0.4 & $0.4$   & $2.27\times10^{-3}$ & $3.84 \times 10^{-4}$ \\
SD23T0.6 & ON & 0.1       &   0.6 & $0.27$   & $8.25\times10^{-4}$ & $1.39 \times 10^{-4}$ \\
SD23T0.8 & ON & 0.1       &   0.8 & $0.2$   & $2.72\times10^{-4}$ & $4.6  \times 10^{-5}$ \\
SD23T1   & ON & 0.1       &   1 & $0.16$     & $9.6 \times10^{-5}$ & $1.6  \times 10^{-5}$ \\
SD23T2   & ON & 0.1       &   2 & $0.08$     & $1.6 \times10^{-5}$ & $2.7  \times 10^{-6}$ \\
SD23T4   & ON & 0.1       &   4 & $0.04$     & $9.7 \times10^{-6}$ & $1.6  \times 10^{-6}$ \\
SD23T6   & ON & 0.1       &   6 & $0.027$     & $1.63\times10^{-6}$ & $2.8  \times 10^{-7}$ \\
SD23T8   & ON & 0.1       &   8 & $0.02$     & $1.43\times10^{-6}$ & $2.4  \times 10^{-7}$ \\


\hline\noalign{\smallskip}
\end{tabular}}

Note: Col. 1: model names. Col.2: the nuclear stars gravity included or not. ``ON" means we include stars gravity. Cols 3, 4: the density and temperature at the outer boundary, respectively. Col.5: Bondi radius in unit of $10 {\rm pc}$. Col. 6: the mass accretion rate measured at the inner boundary of the simulation domain. Col. 7: the mass accretion rate onto the central black hole. Note that the results listed in Cols 6-7 are obtained by time average the data after the simulations achieve a quasi-steady state.

\end{table*}

\begin{table*} \caption{Simulation parameters and results }
\setlength{\tabcolsep}{4mm}{
\begin{tabular}{ccccccc}
\hline \hline
 Model & Stars gravity & $\rho_0$                    & $T_0$  & Bondi radius ($R_B$)    &  $\dot M_{\rm in}(r_{\rm in})$ & $\dot M_{\rm BH}$\\

  &   & ($10^{-22}\text{g cm}^{-3}$)& ($10^7$K) & ($10 {\rm pc}$) & ($L_{\rm Edd}/0.1c^2$)         & ($L_{\rm Edd}/0.1c^2$)   \\
(1) & (2)             & (3)                         &  (4)      &     (5)          &        (6)  & (7)     \\

\hline\noalign{\smallskip}
SD233T0.2& ON & 0.3       &   0.2 & $0.8$   & $2.38\times10^{-2}$ & $4    \times 10^{-3}$ \\
SD233T0.4& ON & 0.3       &   0.4 & $0.4$   & $1.5 \times10^{-2}$ & $2.5  \times 10^{-3}$ \\
SD233T0.6& ON & 0.3       &   0.6 & $0.27$   & $8.9 \times10^{-3}$ & $1.5  \times 10^{-3}$ \\
SD233T0.8& ON & 0.3       &   0.8 & $0.2$   & $5.5 \times10^{-3}$ & $9.3  \times 10^{-4}$ \\
SD233T1  & ON & 0.3       &   1 & $0.16$     & $3.6 \times10^{-3}$ & $6.1  \times 10^{-4}$ \\
SD233T2  & ON & 0.3       &   2 & $0.08$     & $5.6 \times10^{-5}$ & $9.5  \times 10^{-6}$ \\
SD233T4  & ON & 0.3       &   4 & $0.04$     & $1.03\times10^{-5}$ & $1.74 \times 10^{-6}$ \\
SD233T6  & ON & 0.3       &   6 & $0.027$     & $4.67\times10^{-6}$ & $7.89 \times 10^{-7}$ \\
SD233T8  & ON & 0.3       &   8 & $0.02$     & $3.24\times10^{-6}$ & $5.48 \times 10^{-7}$ \\
SD22T0.2 & ON & 1         &   0.2 & $0.8$   & $4.53\times10^{-2}$ & $7.66 \times 10^{-3}$ \\
SD22T0.4 & ON & 1         &   0.4 & $0.4$   & $3.8 \times10^{-2}$ & $6.42 \times 10^{-3}$ \\
SD22T0.6 & ON & 1         &   0.6 & $0.27$   & $3.44\times10^{-2}$ & $5.8  \times 10^{-3}$ \\
SD22T0.8 & ON & 1         &   0.8 & $0.2$   & $3.19\times10^{-2}$ & $5.39 \times 10^{-3}$ \\
SD22T1   & ON & 1         &   1 & $0.16$     & $3.15\times10^{-2}$ & $5.32 \times 10^{-3}$ \\
SD22T1H & ON & 1         &   1 & $0.16$     & $3.2 \times10^{-2}$ & $5.4  \times 10^{-3}$ \\
SD22T2   & ON & 1         &   2 & $0.08$     & $2.2 \times10^{-2}$ & $3.7  \times 10^{-3}$ \\
SD22T4   & ON & 1         &   4 & $0.04$     & $1.3 \times10^{-3}$ & $2.2  \times 10^{-4}$ \\
SD22T6   & ON & 1         &   6 & $0.027$     & $1.94\times10^{-5}$ & $3.3  \times 10^{-6}$ \\
SD22T8   & ON & 1         &   8 & $0.02$     & $1.06\times10^{-5}$ & $1.8  \times 10^{-6}$ \\

\hline\noalign{\smallskip}
\end{tabular}}

Note: Col. 1: model names. Col.2: the nuclear stars gravity included or not. ``ON" means we include stars gravity. Cols 3, 4: the density and temperature at the outer boundary, respectively. Col.5: Bondi radius in unit of $10 {\rm pc}$. Col. 6: the mass accretion rate measured at the inner boundary of the simulation domain. Col. 7: the mass accretion rate onto the central black hole. Note that the results listed in Cols 6-7 are obtained by time average the data after the simulations achieve a quasi-steady state.

\end{table*}

\begin{table*} \caption{Simulation parameters and results }
\setlength{\tabcolsep}{4mm}{
\begin{tabular}{ccccccc}
\hline \hline
 Model & Stars gravity & $\rho_0$                    & $T_0$  & Bondi radius ($R_B$)    &  $\dot M_{\rm in}(r_{\rm in})$ & $\dot M_{\rm BH}$ \\
  &   & ($10^{-22}\text{g cm}^{-3}$)& ($10^7$K) & ($10 {\rm pc}$) & ($L_{\rm Edd}/0.1c^2$)         & ($L_{\rm Edd}/0.1c^2$)  \\
(1) & (2)             & (3)                 &  (4)      &     (5)          &        (6)   & (7)        \\

\hline\noalign{\smallskip}
D24T0.4 & OFF & 0.01      &   0.4& $0.4$   & $2.34 \times10^{-5}$ & $3.96  \times 10^{-6}$ \\
D24T4   & OFF & 0.01      &   4 & $0.04$    & $2.94 \times10^{-7}$ & $4.97  \times 10^{-8}$ \\
D22T4   & OFF & 1         &   4 & $0.04$    & $7    \times10^{-4}$ & $1.2   \times 10^{-4}$ \\

 \hline\noalign{\smallskip}
\end{tabular}}

Note: Col. 1: model names. Col.2: the nuclear stars gravity included or not. ``OFF" means we do not include stars gravity. Cols 3, 4: the density and temperature at the outer boundary, respectively. Col.5: Bondi radius in unit of $10 {\rm pc}$. Col. 6: the mass accretion rate measured at the inner boundary of the simulation domain. Col. 7: the mass accretion rate onto the central black hole. Note that the results listed in Cols 6-7 are obtained by time average the data after the simulations achieve a quasi-steady state.

\end{table*}

\section{Results}
The models with nuclear stars gravity are summarized in Tables 1 and 2. Our purpose is to find a formula to calculate the black hole accretion rate as a function of density and temperature at parsec scale. Therefore, we perform many simulations with different density and temperature at parsec scale.

In column 5 of Tables 1 and 2, we show the Bondi radius (in unit of 10 pc). In Tables 1 and 2, we totally have 77 models. For five models with $T_0=10^6 {\rm K}$, the Bondi radius is 16 pc, which is slightly larger than our outer radial boundary of the simulations. For other 72 models with higher $T_0$, the Bondi radius is much smaller than 10 pc. For these 72 models, the Bondi radius is inside our computational domain.

The black hole mass accretion rate is summarized in Tables 1 and 2. The inner boundary of our simulation domain is $500r_s$. The mass accretion rate at $500r_s$ can be directly obtained in our simulations. A question is that how to calculate the accretion rate at black hole horizon. In this paper, we study low angular momentum accretion. We set the ``circularization radius" ($r_{cir}$) to be 350$r_s$, which is much smaller than the inner boundary of our simulation domain. We assume that when gas crosses the inner boundary, it can freely fall to $r_{cir}$. The mass accretion rate is constant with radius in the region $r_{cir} < r < 500r_s$. When gas arrives at $r_{cir}$, a rotationally supported disk can form. In the presence of viscosity, a viscous accretion flow will form. Wind can be generated in the viscous accretion flow. Due to the presence of wind, the radial distribution of mass inflow rate outside 10$r_s$ can be described as a power law function of radius with a power law index $s$ (Stone et al. 1999; Sadowski et al. 2013). Inside 10$r_s$, the black hole gravity is so strong, no wind can form, so $s=0$ (Sadowski et al. 2013). Stone et al. (1999) perform hydrodynamical (HD) simulations of hot accretion flow with Newtonian potential. They find that $s=0.73$. Stone et al. (1999) assume Newtonian potential, their result can not be applied to the region very close to the black hole (inside 10$r_s$). However, their result can be applied to the region away from the black hole (outside $10r_s$). Sadowski et al. (2013) perform General relativistic simulations of hot accretion flows. They find that outside $10r_s$, s=0.5. Inside $10r_s$, $s=0$. In our previous works (Bu et al. 2013), we also find that outside $10r_s$, s=0.5. Inside $10r_s$, $s=0$. In the present paper, we assume $s=0.5$ outside $10r_s$. Inside $10r_s$, $s=0$. Therefore, the black hole accretion rate can be calculated using the formula:
\begin{equation}
\dot M_{\rm BH}=\dot M_{\rm in}\left(\frac{10r_s}{r_{cir}}\right)^{0.5}
\end{equation}
$\dot M_{\rm in}$ is the mass inflow rate at the inner boundary of the simulation domain.

We also need to note one important issue. It is the calculation of radiative efficiency of the central AGN in our simulations. The radiative efficiency is used to calculate the luminosity of the central AGN. The luminosity is used to calculate the Compton heating, photoionization heating. The exact formula of radiative efficiency of hot accretion flow is given in Xie \& Yuan (2012). The radiative efficiency depends on mass accretion rate, viscosity coefficient $\alpha$ and the parameter $\delta$ which describes the fraction of the direct
viscous heating to electrons. When we calculate the radiative efficiency of the central AGN, we choose the viscosity coefficient $\alpha=0.1$ (Xie \& Yuan 2012). This is because that observations of accretion systems show that $\alpha \sim 0.1$ (King et al. 2007). About the value of $\delta$, for the hot accretion flow in our Galactic center, Yuan et al. (2003) found that $\delta\approx 0.5$. Therefore, in this paper, we use $\delta=0.5$ to calculate the radiative efficiency of the central LLAGN. We refer to Xie \& Yuan (2012) and Bu \& Yang (2018) for the details of calculation of radiative efficiency.

\subsection{Black hole accretion rate}

It is clear that for a given temperature, the black hole accretion rate increases with the increase of density at outer boundary. Bu \& Yang (2018) studied the accretion flow at parsec scale irradiated by a LLAGN without nuclear stars gravity. In Bu \& Yang (2018), it is also found that black hole accretion rate increases with the increase of density at outer boundary. We also give very detailed explanation about the results in Bu \& Yang (2018). For convenience, we also briefly introduce here. Gas heating and cooling have different dependence on density. Gas cooling is more sensitive to density. For example, Compton heating is $\propto \rho$; Bremsstrahlung cooling is $\propto \rho^2$. With increase of density, cooling increases faster than heating. Gas temperature (gas pressure gradient force) will decrease which will result in incrase of infall velocity. Given that infall velocity and density increase, central black hole accretion rate increases.
We find when $\rho_0$ is bigger than $10^{-22} \text{g } {\rm cm}^{-3}$, luminosity of central AGN will be higher than 2$\% L_{Edd}$. Therefore, we do not perform simulations with $\rho_0 > 10^{-22} \text{g } {\rm cm}^{-3}$

If we fix the gas density at the outer boundary, black hole accretion rate will become smaller with increase of temperature at outer boundary. The reason is easy to be understood. With the increase of
temperature, the specific energy of gas increase, gas can more easily escape to form wind. We note that the wind is thermally driven instead of magnetic driven (e.g.,Blandford \& Payne 1982; Emmering et al. 1992; Romanova et al. 1997; Bottorff et al. 2000; Li \& Begelman 2014).

We use a power-law function of density and temperature to fit the black hole mass accretion rate. This is because power-law function is more suitable for accretion theory. For example, the physical variables of black hole hot accretion flow can be well described by power-law function of radius (Narayan \& Yi 1994). We find that the black hole mass accretion rate can be described as
\begin{equation}
\frac{\dot M_{\rm BH} }{\dot M_{\rm Edd}}=10^{-3.11} \left( {\frac{\rho_0}{10^{-22}\rm{g\cdot cm^{-3}}}} \right)^{1.36} \left({\frac{T_0}{10^7\rm K}}\right)^{-1.9}
\label{mdot}
\end{equation}
$\rho_0$ and $T_0$ are density and temperature at outer boundary (10pc).
In Tables 1 and 2, we have totally 77 models. For 58 models (mainly with density in the range $10^{-26}\text{g cm}^{-3} < \rho < 10^{-23}\text{g cm}^{-3} $), the predicted accretion rate by Equation (\ref{mdot}) deviates from the time-averaged simulation result by a factor smaller than 3. For 70 models, the predicted accretion rate by Equation (\ref{mdot}) deviates from the time-averaged simulation result by a factor smaller than 5. For the other 7 models, the predicted accretion rate by Equation (\ref{mdot}) deviates from the time-averaged simulation result by a factor smaller than $\sim 10$. Therefore, Equation (\ref{mdot}) is good enough to represent the simulation results. If other complex function is used to fit the black hole accretion rate, we can obtain formula with very tiny residual error. However, the physical meanings of complex function can not be straightforward.

In this paper, we find $\dot M_{\rm BH} \propto \rho_0^{1.36} T_0^{-1.9}$. Bondi solution predicts that $\dot M_{\rm BH} \propto \rho_0 T_0^{-1.5}$.  Now, we explain why the accretion rate dependence found in this paper is different from that of Bondi accretion model. Bondi accretion model predicts that $\dot M_{\rm BH} \propto \rho_0 T_0^{-1.5} \propto \rho_0 R_B^2 c_s$. $R_B$ and $c_s$ are Bondi radius and sound speed at Bondi radius, respectively. At Bondi radius, Bondi formula predicts infall velocity is roughly equal to sound speed. Bondi solution with $\gamma=5/3$ is calculated without radiative cooling. Non-radiative hydrodynamic equations are density free. We take hot accretion flow as an example to explain what is ``density free". For hot accretion flows, radiative cooling can be neglected. Suppose that there are two hot accretion flow solutions. The first solution has higher mass accretion rate (or density) than the second solution. The absolute values of temperature and velocity at any radii in the first solution are same as those at the same radii in the second solution. One example is the self-similar solution of hot accretion flow (e.g., Narayan \& Yi 1994). In the self-similar solution, the temperature and velocity do not change with mass accretion rate (or density). For the Bondi solution with $\gamma=5/3$, the change of density can not affect other properties of gas (e.g., velocity, temperature); Also, the change of gas temperature (or sound speed) can not change density. In the simulations in this paper, we have cooling and heating. The change of density will also change other properties of gas. For example, as introduced above, the change of density can change the radial infall velocity. With the increase of outer boundary density, the infall velocity also increases (see also Yang \& Bu 2018). Therefore, in Equation (\ref{mdot}), the power law index of density would be larger than unity because radiative cooling reduces the pressure gradient around the Bondi radius. In Equation (\ref{mdot}), the power law index of temperature is -1.9. This means that for high temperature models, the accretion rate deviation from the Bondi value is larger compared to the low temperature model. The reason is as follows. Almost in all models with an initial high temperature (equal or higher than $4\times 10^7$ K), the black hole accretion rate is very low (see Figure \ref{Fig:mdot2} as an example). As introduced below, when accretion rate is low, Compton heating is not important. Compression work heating may be the reason for the wind generation. In this case, we can expect the higher the temperature of gas, the easier the wind to be generated. Therefore, with the increase of temperature, the black hole accretion rate will deviate more from the Bondi accretion rate.

\begin{figure}
\begin{center}
\includegraphics[scale=0.5]{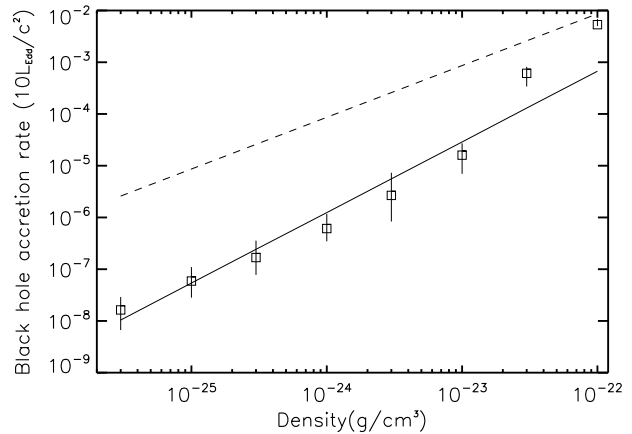}\hspace*{0.7cm}
\hspace*{0.5cm} \caption{Black hole mass accretion rate in unit of $\dot M_{\rm Edd}$ versus density at outer boundary (or infinity) with temperature at outer boundary (infinity) fixed to be $T_0=10^7 {\rm K}$. The solid line is from the fitting formula (Equation (\ref{mdot})). Squares correspond to the time-averaged values of mass accretion rate of simulations listed in Tables 1 and 2. The error bars over-plotted on squares represent the change range of simulations due to fluctuations. The dashed line is calculated according to Bondi formula by assuming $\gamma=5/3$. \label{Fig:mdot}}
\end{center}
\end{figure}

\begin{figure}
\begin{center}
\includegraphics[scale=0.45]{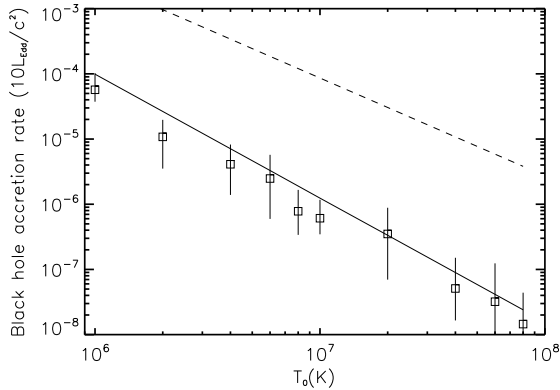}\hspace*{0.7cm}
\hspace*{0.4cm} \caption{Black hole mass accretion rate in unit of $\dot M_{\rm Edd}$ versus temperature at outer boundary (or infinity) with density at outer boundary (infinity) fixed to be $\rho_0=10^{-24} {\rm g/cm^3}$. The solid line is from the fitting formula (Equation (\ref{mdot})). Squares correspond to the time-averaged values of mass accretion rate of simulations listed in Tables 1 and 2. The error bars over-plotted on squares represent the change range of simulations due to fluctuations. The dashed line is calculated according to Bondi formula by assuming $\gamma=5/3$. \label{Fig:mdot2}}
\end{center}
\end{figure}

In Figure 1, we plot the black hole mass accretion rate in unit of $\dot M_{\rm Edd}$ versus density at outer boundary. In this plot, we fix the temperature at outer boundary (infinity) to be $T_0=10^7 {\rm K}$. The solid line is from the fitting formula (Equation (\ref{mdot})). Squares correspond to the time-averaged values of mass accretion rate of simulations listed in Tables 1 and 2. The error bars over-plotted on squares represent the change range of simulations due to fluctuations. The dashed line is calculated according to Bondi formula by assuming $\gamma=5/3$. For $\rho_0 \leq 10^{-23} \rm g /\rm {cm}^3$, the fitting formula can well represent the simulation result. For the model with $\rho_0 = 3\times 10^{-23} \rm g /\rm {cm}^3$, the fitting formula deviates from the simulation result (lower edge of error bar) by a factor smaller than 3.
For the model with $\rho_0 = 10^{-22} \rm g /\rm {cm}^3$, the fitting formula deviates from the simulation result (lower edge of error bar) by a factor $\sim 10$. Therefore, the fitting formula is not accurate when density is close to $10^{-22} \rm g /\rm {cm}^3$. But for $\rho_0 \leq 3 \times 10^{-23} \rm g /\rm {cm}^3$, the fitting formula is good enough to represent the simulation results.
The Bondi accretion rate depends on $\gamma$ which varies from 1 to $5/3$. If the density and temperature at Bondi radius are fixed, smallest Bondi accretion rate corresponds to $\gamma=5/3$. In all figures in this paper, the Bondi accretion rate is calculated by assuming $\gamma=5/3$. The predicted accretion rate by Bondi formula can be 2 orders of magnitude higher than that of simulations. The reason is that Bondi formula assumes no wind is present; all the gas captured at outer boundary can fall onto the black hole. However, as noted above, in reality wind is present which reduces the black hole accretion rate significantly. Note that if we assume $\gamma$ is smaller than 5/3, we will get even larger Bondi accretion rate.

In Figure 2, we plot the black hole mass accretion rate in unit of $\dot M_{\rm Edd}$ versus temperature at outer boundary. In this plot, we fix the density at outer boundary (infinity) to be $\rho_0=10^{-24} {\rm g/cm^3}$. The solid line is from the fitting formula (Equation (\ref{mdot})). Squares correspond to the time-averaged values of mass accretion rate of simulations listed in Tables 1 and 2. The error bars over-plotted on squares represent the change range of simulations due to fluctuations. The dashed line is calculated according to Bondi formula by assuming $\gamma=5/3$. Again, we find the fitting formula can well represent the simulation result. The Bondi formula again over estimates the black hole accretion rate by 2 orders of magnitude.

\subsection{The reason for the deviation of black hole accretion rate from the Bondi rate}

\begin{figure*}
\begin{center}
\includegraphics[scale=0.45]{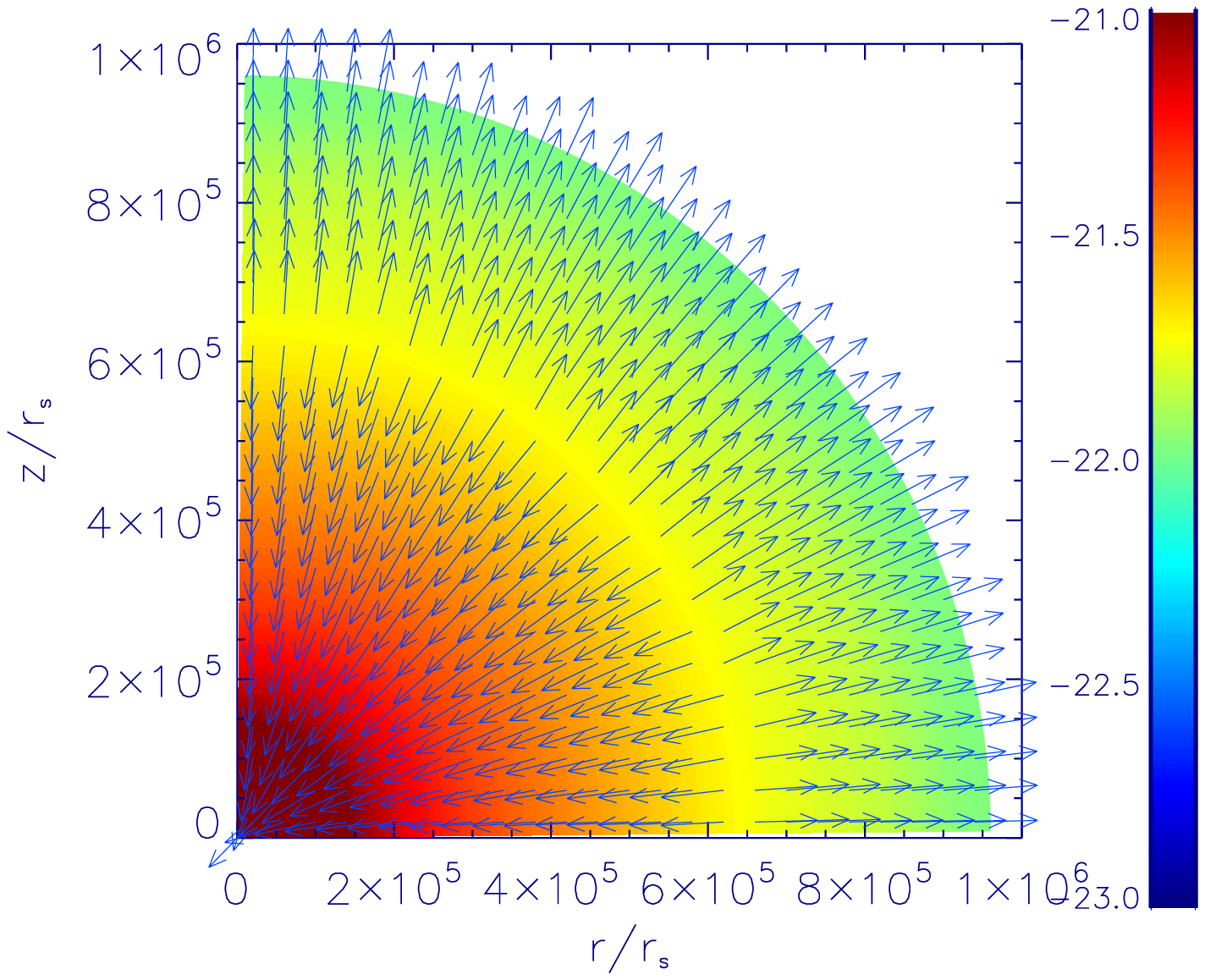}\hspace*{0.7cm}
\includegraphics[scale=0.45]{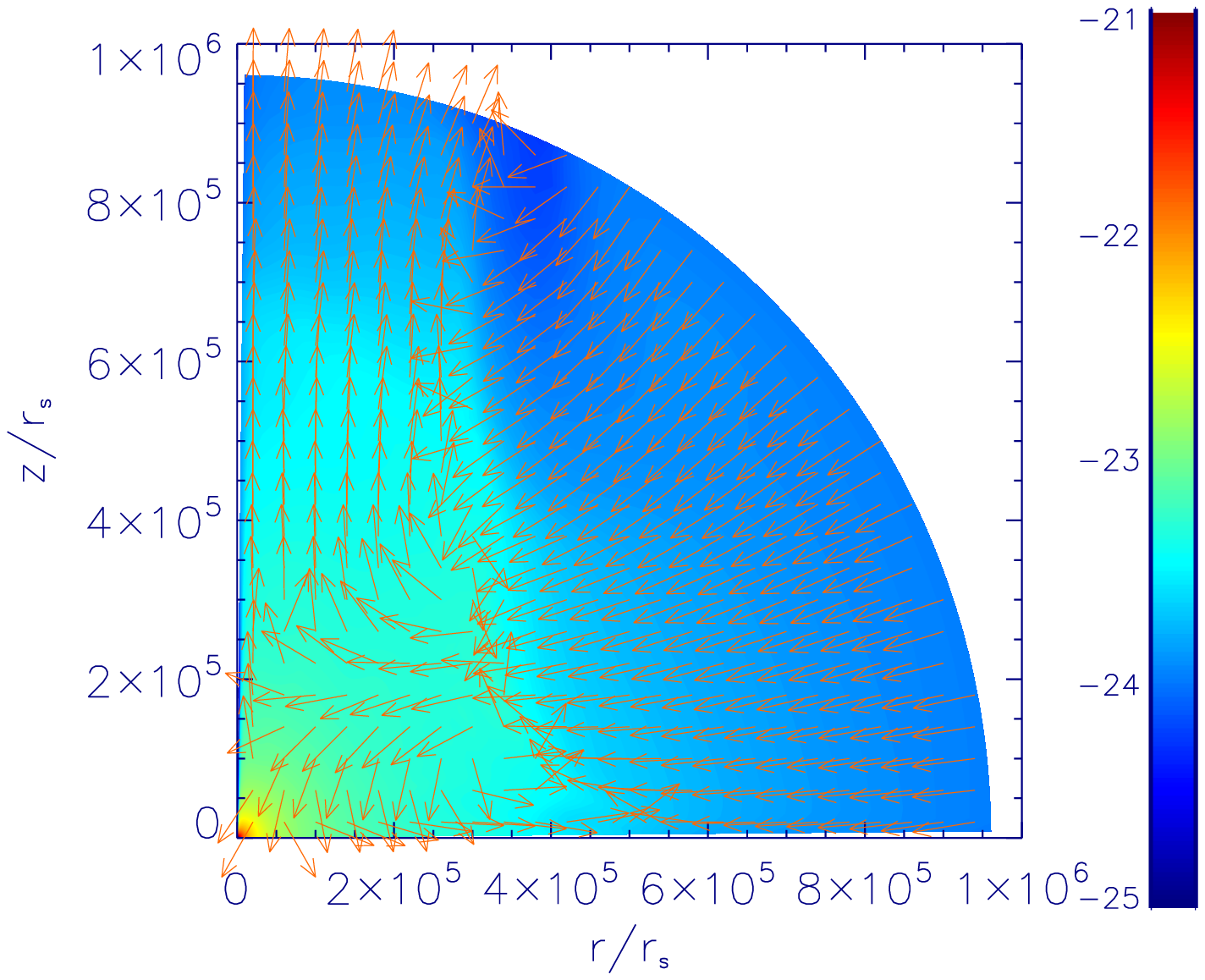}\hspace*{0.7cm}
\hspace*{0.5cm} \caption{Snapshot ($t=6.6\times 10^5$ year) of the accretion flow when wind is present. Left panel is for model SD22T1. Right panel is for model SD24T1. Colors shows logarithm density. Arrows show the unit velocity vector. \label{Fig:snapshot}}
\end{center}
\end{figure*}

\begin{figure*}
\begin{center}
\includegraphics[scale=0.45]{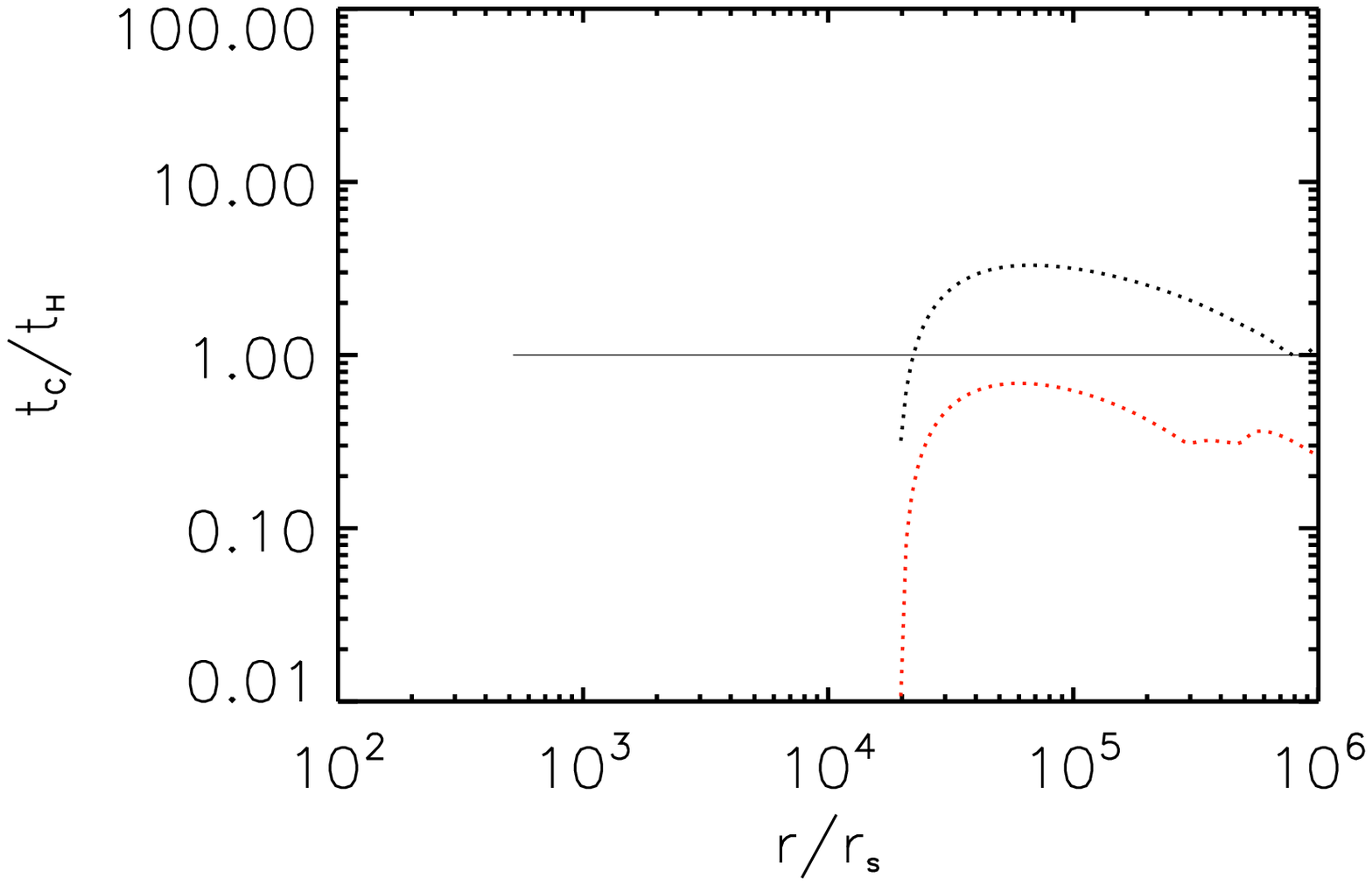}\hspace*{0.7cm}
\includegraphics[scale=0.45]{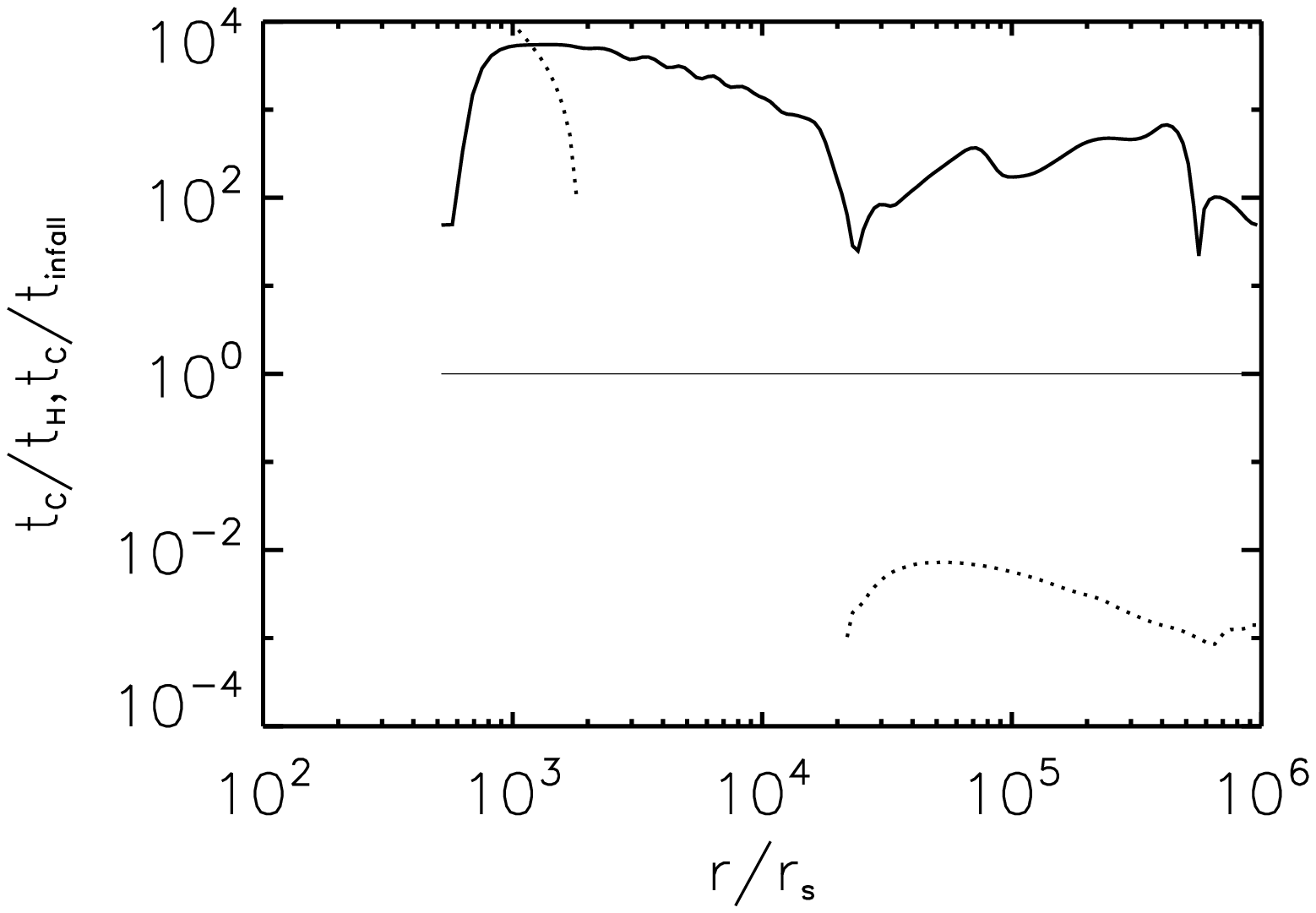}\hspace*{0.7cm}
\hspace*{0.5cm} \caption{Dotted lines in the two panels show the radial profile of cooling timescale over heating timescale. The heating timescale is calculated by using Compton heating. The cooling timescale is calculated by using bremsstrahlung cooling. The left panel is for model SD22T1. When plotting the dotted lines this panel, we fixed $\theta=45^\circ$. Because the flow is more spherical, the result at other theta angle is almost same. The black dotted line is for the snapshot at $t=6.6\times10^5$ year when wind is present. The red dotted line is for the snapshot at $t=7\times10^5$ year when wind is absent. The right panel is for model SD24T1. When plotting the dotted line this panel, we fix $\theta=8^\circ$. At this angle, wind is present. We note that for the dotted line in the left pane in the region $r<2\times 10^4 r_s$, there is no data. The reason is that in this region, Compton process plays a cooling role. In the right panel, for the dotted line the absence of data in the region $2\times 10^3r_s<r<2\times 10^4r_s$ is also because in this region Compton process plays a cooling role. The solid line in the right panel shows the ratio of cooling timescale to the radial infall timescale ($r/\mid v_r \mid $ ) around the equatorial plane. \label{Fig:timescale}}
\end{center}
\end{figure*}

We note that the significant reduction of accretion rate from the Bondi rate is due to the presence of wind. In this paper, the gas density at Bondi radius is set to be in the range from $3\times 10^{-26} {\rm g/cm^3} $ to $ 10^{-22} {\rm g/cm^3} $. Therefore, the resulting black hole accretion rate can be very different from model to model. Here, we take models SD22T1 and SD24T1 as two examples to explain the different wind properties in both high and low accretion rate flows. In Figure \ref{Fig:snapshot}, we plot the snapshot of the two models. From this figure, we can see that for the high accretion rate model (SD22T1), the wind is much spherical. In Bu \& Yang (2018), we only study the flow with relative high accretion rate. We also find that the wind is nearly spherical. In the left panel of Figure \ref{Fig:timescale}, we plot the radial profile of cooling timescale over heating timescale when wind is present at $t=6.6\times 10^5$ year (black dotted line). We can see that in the wind region, Compton heating timescale is shorter than radiative cooling timescale. Therefore, the wind is Compton heated. Due to the Compton heating, wind becomes energetic enough to escape from black hole gravitational potential. The Compton heating rate is spherical. Therefore, wind in this model is also much spherical.

\begin{figure}
\begin{center}
\includegraphics[scale=0.45]{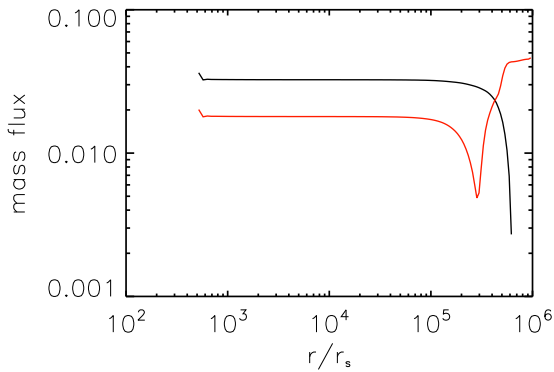}\hspace*{0.5cm}
\caption{Snapshot of the radial profiles of mass inflow
rate for model SD22T1. The black line is for the snapshot at $t=6.6\times 10^5$ year when wind is present. The red line is for the snapshot at $t=7\times 10^5$ year when wind is absent. \label{Fig:mdot-snap}}
\end{center}
\end{figure}

We note that in the high accretion rate model (SD22T1), wind is generated episodically. The reason is as follows.
Initially, there is no wind and the black hole accretion rate is zero. When gas falls to the black hole, the black hole accretion rate (or luminosity) increases. When the luminosity is high enough, the Compton heating will become important. For example, at $t=6.6\times 10^5$ year, the Compton heating timescale can be much smaller than radiative cooling timescale at the outer region (see the black dotted line in the Left panel of Figure \ref{Fig:timescale}). Then gas in the outer region can form wind and escape. When wind takes away gas, the mass inflow rate will become smaller (see the red line for gas inflow rate in Figure \ref{Fig:mdot-snap}). Then black hole luminosity will become smaller and the Compton heating becomes unimportant (see the red dotted line in left panel of Figure \ref{Fig:timescale}). In this case, cooling dominates heating. The inflow rate keeps increasing with time when cooling dominates heating. When the accretion rate becomes high again, the wind will be generated again by Compton heating.

Now we study why wind is present in model SD24T1. The dotted line in the right panel of Figure \ref{Fig:timescale} shows the radial profile of radiative cooling timescale over Compton heating scale at the wind region. Compton heating is not important compared to radiative cooling. Therefore, in this model. The wind is not due to Compton heating.
For the low accretion rate model (SD24T1), we check the ratio of cooling timescale to gas infall timescale ($r/v_r$). We plot the result in the right panel of Figure \ref{Fig:timescale} (solid line). The cooling timescale in this model is much longer than the infall velocity. Therefore, cooling is not important in this model.
The injected gas temperature is $10^7 {\rm K}$ at outer boundary ($10^6r_s$). Correspondingly, the Bondi radius is at $3.2\times10^5r_s$. Therefore, at the outer boundary, the gas internal energy is higher than its gravitational energy. In other words, the Bernoulli parameter of gas is positive. Gas is unbound. When gas falls in, the compression work also heats gas. We find that at the wind region, gas pressure gradient force is larger than gravitational force.

\begin{figure*}
\begin{center}
\includegraphics[scale=0.45]{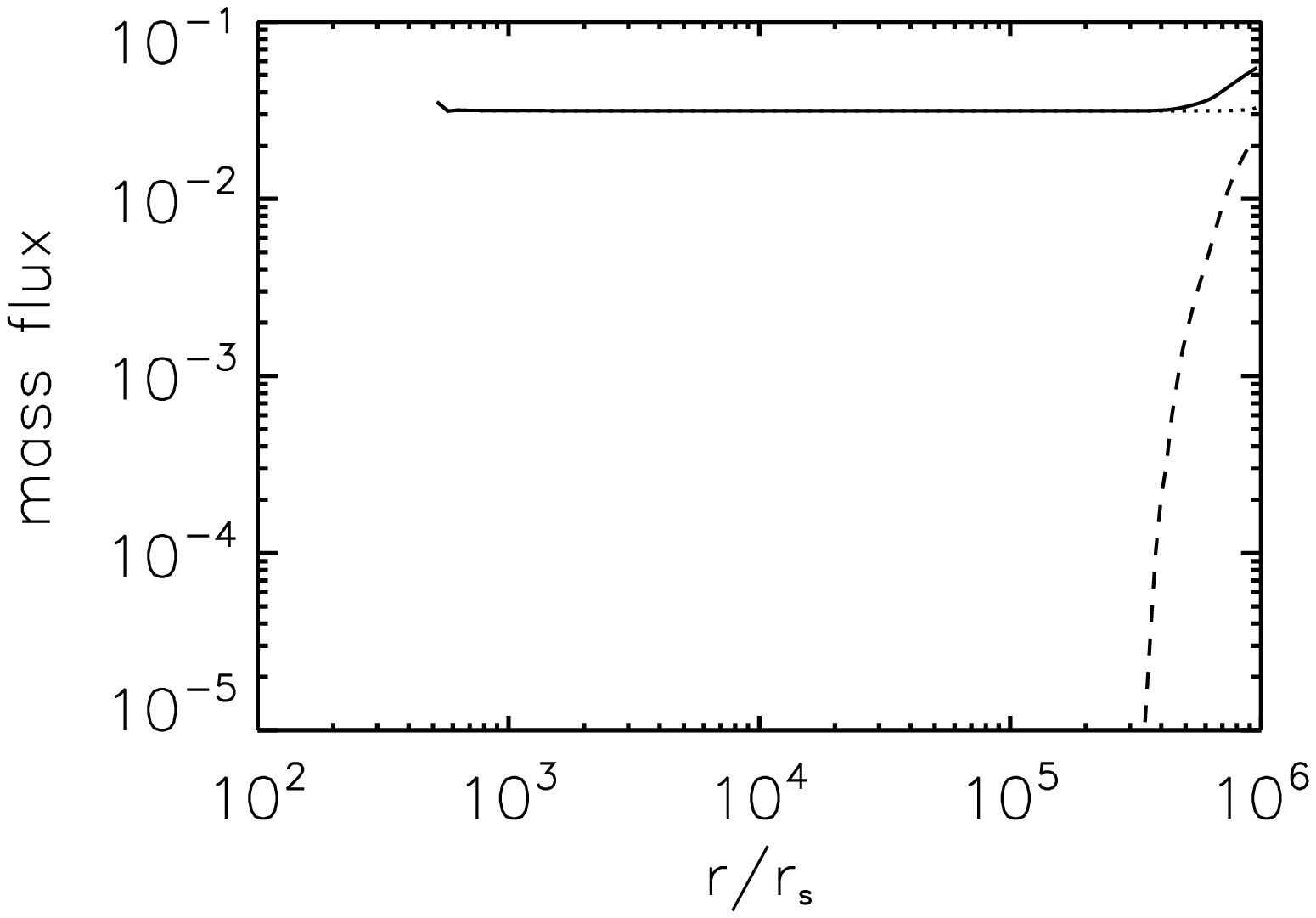}\hspace*{0.5cm}
\includegraphics[scale=0.45]{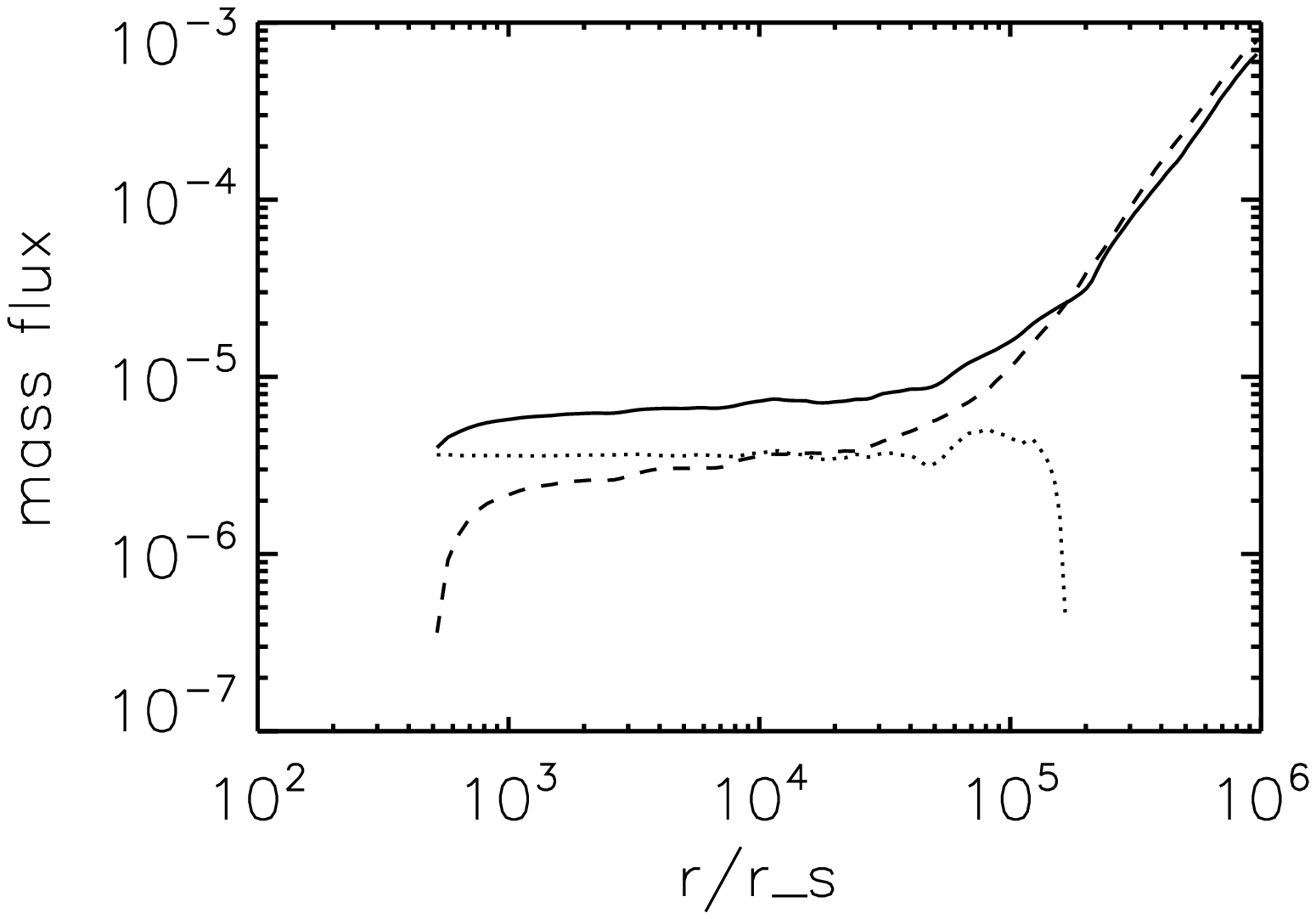}\hspace*{0.5cm}
\caption{The radial profiles of time-averaged (from $t=1.5 \times
10^5$ to $1.15 \times 10^6$ year) and angle integrated mass inflow
rate (solid line), outflow rate (dashed line) and the net rate
(dotted line). The left panel is for model SD22T1. The right panel is for model SD24T1. The mass fluxes are expressed in unit of the
Eddington accretion rate.\label{Fig:mdota}}
\end{center}
\end{figure*}

To quantitatively study the properties of the inflow/outflow
component, we calculate the radial dependence of mass inflow,
outflow, and net rates as follows: (1) inflow rate
\begin{equation}
\dot {M}_{\rm in} (r)=4\pi r^2 \int_{\rm 0}^{\rm \pi/2}
\rho (r, \theta) \min \left(v_r(r,\theta), 0 \right) \sin\theta d\theta
\end{equation}
(2) outflow rate
\begin{equation}
\dot {M}_{\rm out} (r)=4\pi r^2 \int_{\rm 0}^{\rm \pi/2}
\rho (r, \theta) \max \left(v_r(r,\theta), 0 \right) \sin\theta d\theta
\end{equation}
and (3) net rate
\begin{equation}
\dot {M}_{\rm net} (r)=4\pi r^2 \int_{\rm 0}^{\rm \pi/2}
\rho (r, \theta) v_r(r,\theta) \sin\theta d\theta
\end{equation}
When we calculate the mass inflow rate in Equation (6), we need to pick out inflowing gas. Therefore, in this equation, we have $\min(v_r(r,\theta),0)$. When we calculate the mass outflow rate in Equation (7), we need to pick out outflowing gas. Therefore, in this equation, we have $\max(v_r(r,\theta),0)$.
In Figure \ref{Fig:mdota}, we plot the result. In model SD22T1, wind is present outside $6 \times 10^5 r_s$ (see the velocity vector in left panel of Figure \ref{Fig:snapshot}). Inside $6 \times 10^5 r_s$, wind is absent. Correspondingly, from the left panel of (\ref{Fig:mdota}), we can see that the mass inflow rate just keeps decreasing from outer boundary to $6 \times 10^5 r_s$. Because, $6 \times 10^5 r_s$ is much closer to the outer boundary, the mass accretion rate does not decrease much from outer boundary to $6 \times 10^5 r_s$. This is the reason why in Figure \ref{Fig:mdot}, when outer boundary density is close to $10^{-22} {\rm g cm^{-3}}$, the mass accretion rate does not differ much from the Bondi value. In model SD22T1, Compton heating provides winds and drives outflows at $r>2\times 10^4 r_s$, but the mass outflow rate is not significant. The outflow actually does not affect the inflow and accretion rate. Thus, the profile shown in Figure \ref{Fig:mdota} is constant at Bondi accretion rate. In model SD24T1, wind is present in the region outside $5 \times 10^4 r_s$ (see the velocity vector along equatorial plane in right panel of Figure \ref{Fig:snapshot}). Therefore, the mass inflow rate keeps decreasing from the outer boundary to $5 \times 10^4 r_s$ (see right panel of Figure \ref{Fig:mdota}). The mass inflow rate at the inner boundary is more than 2 orders of magnitude smaller than that at outer boundary.

\subsection{Comparison to Observations}

Now, we check whether the fitting formula (Equation (\ref{mdot})) is reliable. Using Equation (\ref{mdot}), we can predict black hole accretion rate and calculate luminosity of accretion flow (see Section 2.2). We can compare the predicted luminosity by Equation (\ref{mdot}) to observations of LLAGNs. In order to do the comparison, we need to look for observations of LLAGNs. The LLAGNs must satisfy three conditions. First, black hole mass of LLAGNs is $\sim 10^8 M_\odot$; Second, gas density and temperature around the Bondi radius of these LLAGNs are known; Finally, luminosity of these LLAGNs has been given by observations. We have three LLAGNs in literatures. They are NGC 4438 (Machacek et al. 2004), NGC 1316 (Tremaine et al. 2002; McElroy 1995; Kim \& Fabbiano 2003) and NGC 1291 (Tremaine et al. 2002; McElroy 1995; Irwin et al. 2002). We also refer to Pellegrini (2005) for the properties of these LLAGNs. The temperature of gas around Bondi radius of these three LLAGNs is in the range $3.9-7.2\times 10^6 {\rm K}$. We show the results in Figure \ref{Fig:luminosity}. In this figure, the two solid lines are calculated by using the fitting formula (Equation (\ref{mdot})). The upper and lower solid lines correspond to $T_0=3.9\times 10^6 {\rm K}$ and $7.2\times 10^6 {\rm K}$, respectively. The upper and lower dashed lines are calculated according to Bondi formula by assuming $T_0=3.9\times 10^6 {\rm K}$ and $7.2\times 10^6 {\rm K}$, respectively. The square, triangle and diamond represent NGC 4438, NGC1316 and NGC 1291, respectively. The Bondi formula over estimates the luminosity of these three LLAGNs by 2 orders of magnitude. The fitting formula (Equation (\ref{mdot})) can accurately predict the luminosity of these three LLAGNs.

\subsection{Comparison to previous works}

Narayan \& Yi (1994) expected that hot accretion flow is convectively unstable, because entropy of the flow increases towards the black hole. Convection dominated accretion flow model (CDAF) was developed ten years ago (Quataert \& Gruzinov 2000; Narayan et al. 2000). CDAF model assumes that convection transports angular momentum inward. If viscosity is weak, the outward angular momentum transfer by viscosity can be balanced by inward angular momentum transfer by convection. The net accretion rate onto the black hole can be negligibly smaller than the Bondi rate. The inflowing gas is locked in convective eddies.

We take model SD22T1 as an example to compare the results in our paper to CDAF model. In model SD22T1, $\rho_0=10^{-22} {\rm g/cm^3}$, $T_0=10^7 {\rm K}$. Solid lines in Figure \ref{Fig:density22T1} are the time and angle averaged density (left panel) and temperature (right panel) for model SD22T1. In the left panel, dashed line corresponds to $\rho \propto r^{-1/2}$, which corresponds to a CDAF solution. The dotted line is for a density profile $\rho \propto r^{-3/2}$. From this figure, we can see, the density profile differs significantly from that of a CDAF solution. The reason may be as follows. In this paper, we study low angular momentum accretion. ``Circularization" radius is much smaller than the inner boundary of the simulation domain. The flow is much similar to spherical accretion. Therefore, the density profile is much similar to spherical Bondi solution. In the CDAF model, the gas angular momentum is important. In CDAF model, gas is rotationally supported and the gas density is $\rho \propto r^{-1/2}$. The right panel shows temperature. The dashed line in this panel shows the Virial temperature. For $r> 10^4r_s$, gas temperature is much smaller than the Compton temperature of central X-ray photons ($10^8 {\rm K}$). Therefore, in this region, gas is Compton heated. Temperature of gas in this region is higher than Virial temperature. This is the reason for the presence of thermal wind. In the region $r< 10^4r_s$, Compton scaterring is a cooling term. Therefore, in this region, the temperature of gas is much smaller than Virial temperature.

\begin{figure}
\begin{center}
\includegraphics[scale=0.5]{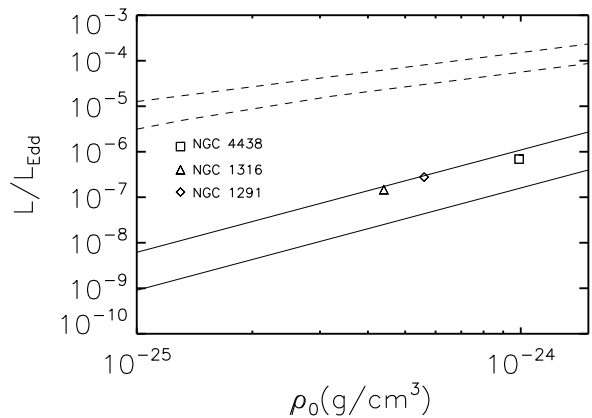}\hspace*{0.7cm}
\hspace*{0.5cm} \caption{Luminosity of accretion flow in unit of $L_{\rm Edd}$ versus density at outer boundary (or infinity). The two solid lines are calculated by using the fitting formula (Equation (\ref{mdot})). The upper and lower solid lines correspond to $T_0=3.9\times 10^6 {\rm K}$ and $7.2\times 10^6 {\rm K}$, respectively. The upper and lower dashed lines are calculated according to Bondi formula by assuming $T_0=3.9\times 10^6 {\rm K}$ and $7.2\times 10^6 {\rm K}$, respectively. The square, triangle and diamond represent NGC 4438, NGC1316 and NGC 1291, respectively. \label{Fig:luminosity}
\label{Fig:luminosity}}
\end{center}
\end{figure}

From right panel of Figure \ref{Fig:snapshot}, we can see the flow in model SD24T1 is not spherical. There is a question. Is the flow CDAF like or spherical accretion like? We plot the time-averaged density profile in Figure \ref{Fig:density24T1}. From this figure, we can see, the density profile of the flow is also more spherical like.

There are lots of previous works on spherical accretion including radiative processes (e.g. Park 1990; Park \& Ostriker 2001). We note these works are analytical works. These works self-consistently include radiative effects. In these works, self-consistent steady solutions are obtained. The solutions found in these papers are time steady. Also, the accretion rate obtained is a constant with radius. In other word, in their solutions, no wind can be found. In our present work, we find that the flows are more like spherical accretion. However, for the high accretion rate model (e.g., SD22T1), the flow is not time steady; winds can be generated episodically. For the low accretion rate model (e.g., SD24T1), wind is always present. Due to the presence of wind, the accretion rate deviates from the Bondi value significantly. This is a big difference between the present work and previous works. More importantly, we give a formula to calculate the black hole accretion rate as a function of density and temperature at Bondi radius. This is a very new point.

\subsection{Test simulations}

In order to study whether the results depend on resolution. We carry out model SD22T1H. The initial conditions and properties of injected gas in model SD22T1H are same as those in model SD22T1. The only difference between models SD22T1H and SD22T1 lies in resolution. Resolution in model SD22T1H is two times that in model SD22T1. The results of SD22T1H are listed in Table 2. Comparing the two models, we find that the results in the two models do not differ much. Our resolution in this paper can capture the main physics of the accretion flows.

We also perform some simulations without nuclear stars gravity shown in Table 3. The purpose is to study the effects of nuclear stars gravity on the results. Each model in Table 3 has its counterpart in Tables 1 or 2. Model D24T0.4 in table 3 has its counterpart SD24T0.4 in table 1. Model D24T4 has its counterpart SD24T4. D22T4 has its counterpart SD22T4. The only difference between models in Table 3 differ from their counterparts in Tables 1 and 2 lies in the nuclear stars gravity. In all the models in Table 3, the nuclear stars gravity is not taken into account. Comparing these models, we can see the results of models without stars gravity differ from those with stars gravity by a factor smaller than 2. The reason is that in the whole computational domain, the black hole gravity dominates.

\begin{figure*}
\begin{center}
\includegraphics[scale=0.45]{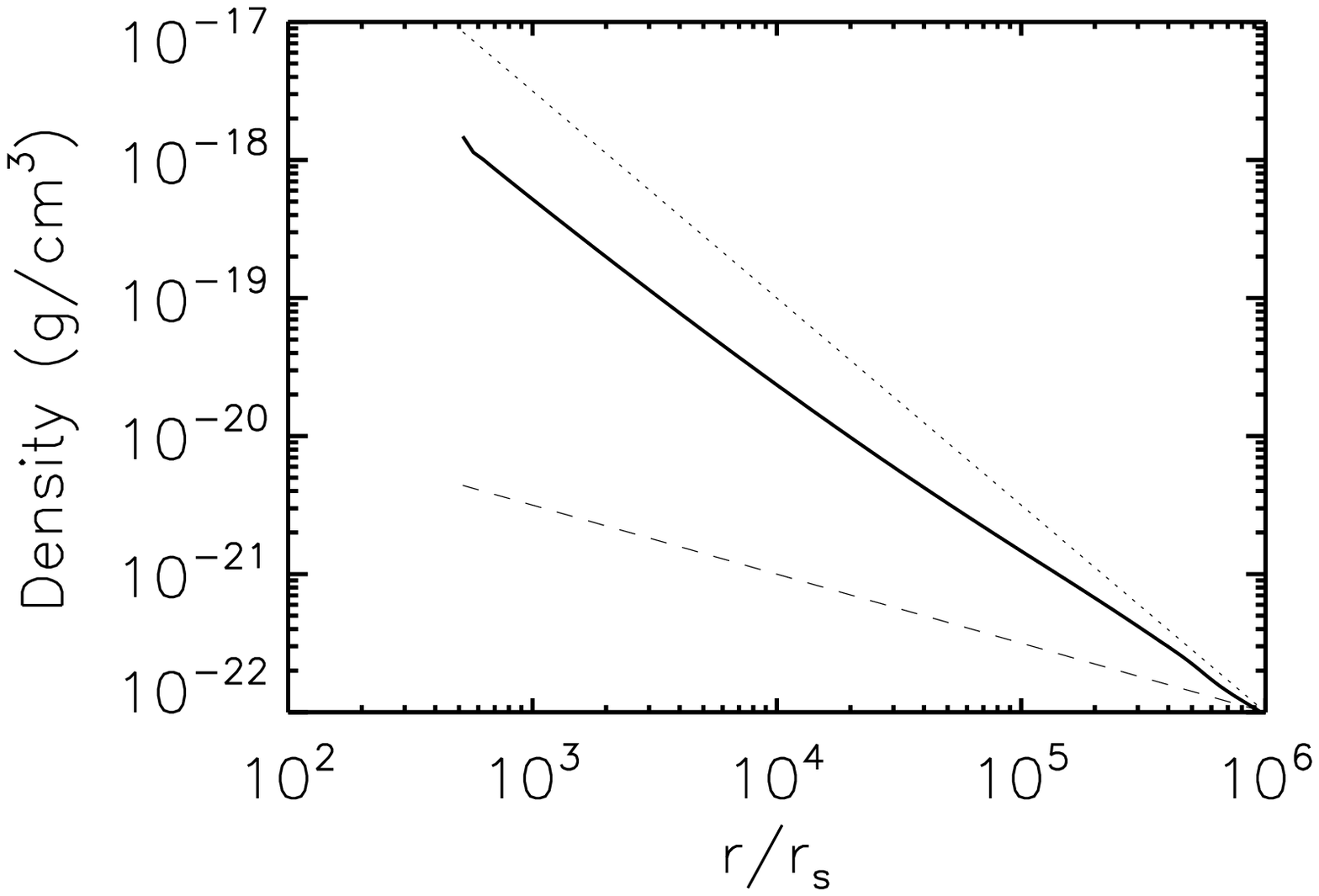}\hspace*{0.7cm}
\includegraphics[scale=0.45]{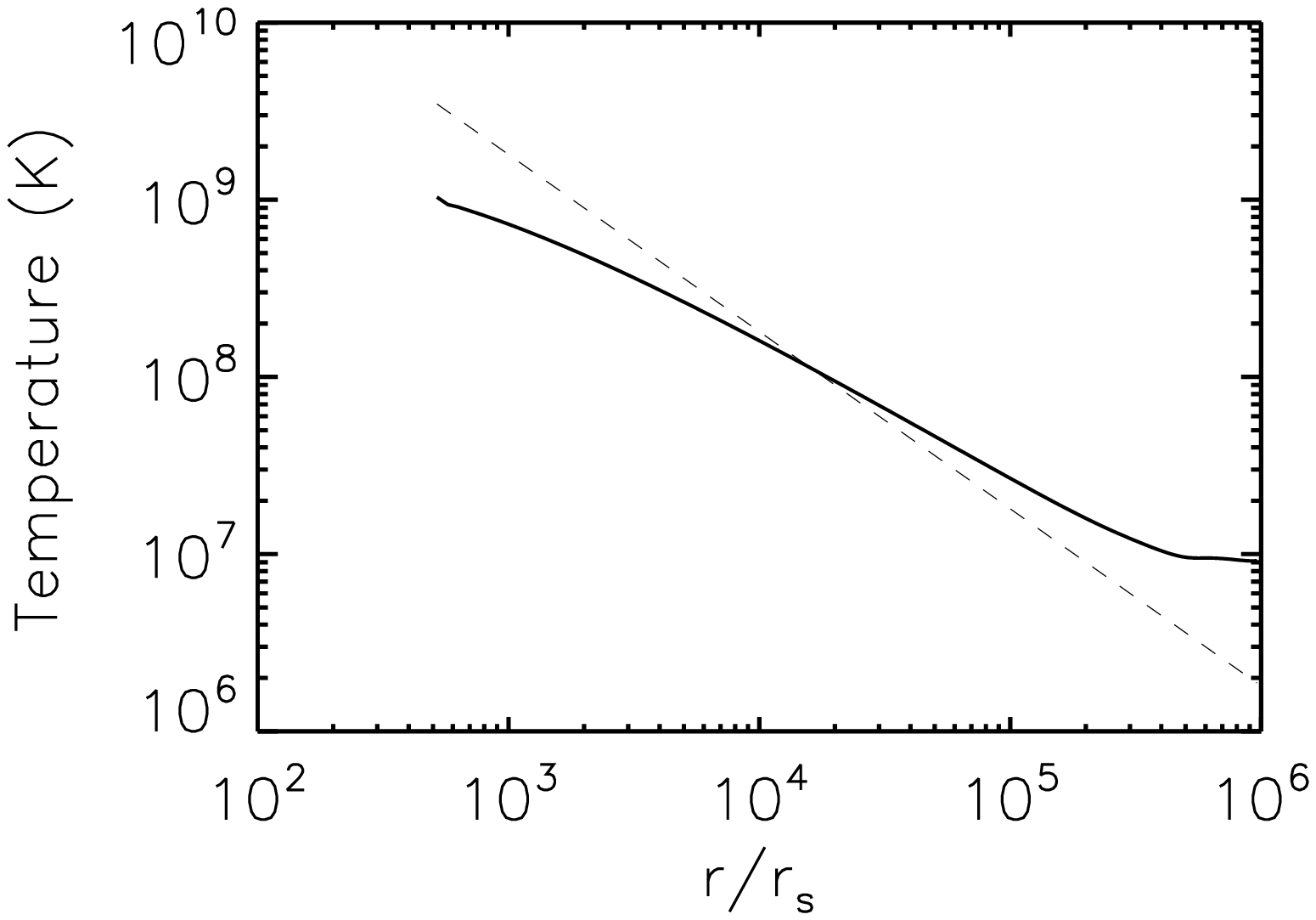}\hspace*{0.7cm}
\hspace*{0.5cm} \caption{Solid lines are the time and angle averaged density (left panel) and temperature (right panel) for model SD22T1. In the left panel, dashed line corresponds to $\rho \propto r^{-1/2}$, which corresponds to a CDAF solution. The dotted line is for a density profile $\rho \propto r^{-3/2}$. The right panel shows temperature. The dashed line in this panel shows the Virial temperature. \label{Fig:density22T1}}
\end{center}
\end{figure*}

\section{Summary and discussion}
In this paper, two-dimensional simulations are performed to investigate slowly rotating accretion flow at parsec and sub-parsec scale irradiated by a LLAGN. Our purpose is to find a formula to calculate the black hole accretion rate based on the density and temperature of gas at parsec scale. In this paper, we set the black hole mass $M=10^8M_{\odot}$.  We obtain formula to calculate black hole accretion rate based on density and temperature of gas at parsec scale. We find the formula can predict the luminosity of LLAGNs (with black hole mass $\sim 10^8M_{\odot}$) very accurately. This formula can be used in the sub-grid models in large scale cosmological simulations (e.g. Springel et al. 2005) with a black hole mass of $\sim 10^8M_{\odot}$.

We note Equations (\ref{mdot}) can be used when $10^{-26}\text{g cm}^{-3} < \rho_0 < 10^{-22}\text{g cm}^{-3} $, $10^6 {\rm K}< T_0 < 10^8 {\rm K}$ and black hole mass $\sim 10^8M_{\odot}$. Properties of radiative accretion flow depend on black hole mass. The formula (\ref{mdot}) can be used for a black hole with mass of $\sim 10^8M_{\odot}$. In future, it is necessary to perform simulations with other values of black hole masses. It will be very useful if a formula (predicting black hole accretion rate) related to black hole mass can be given in future.

\begin{figure}
\begin{center}
\includegraphics[scale=0.5]{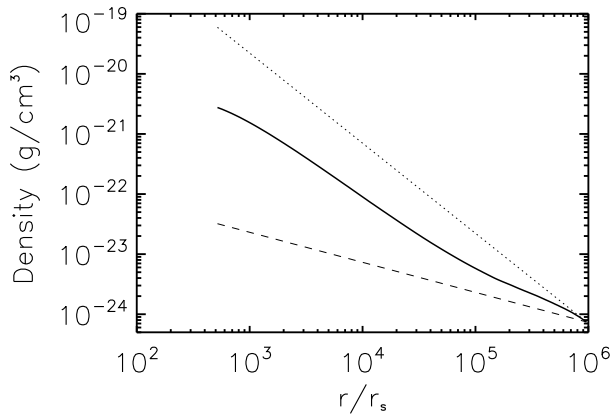}\hspace*{0.7cm}
\hspace*{0.5cm} \caption{As the left panel of Figure \ref{Fig:density22T1}, but for model SD24T1. \label{Fig:density24T1}}
\end{center}
\end{figure}

``Circularization" radius of injected gas is located at 350$r_s$. We have done some test simulations with different values for ``circularization" radius. We find that if the ``circularization" radius is smaller than the inner boundary of the simulation domain, the results do not change much (see also Yang \& Bu 2018). The angular momentum of gas is very small. For the high accretion rate flows (for example model SD22T1), both the time-averaged physical variables (like density shown in left panel of Figure \ref{Fig:density22T1}) and the wind (left panel of Figure \ref{Fig:snapshot}) are very spherical like. For the low accretion rate flows (for example model SD24T1), the time-averaged physical variables (like density shown in Figure \ref{Fig:density24T1}) is much spherical. However, the velocity field is not spherical like (right panel of Figure \ref{Fig:snapshot}). The wind escapes from the region close to the rotational axis. The non-spherical structure of wind is due to the presence of angular momentum. The opening angle of wind may be an important parameter in AGN feedback study, it may determine the interaction efficiency between wind and ISM. In future it is necessary to study high angular momentum flows.

\section*{Acknowledgments}
We thank Z. Yan for useful discussions. This work is supported in part by the Natural Science Foundation of China (grants 11773053,  11573051, 11633006 and 11661161012), the Natural Science
Foundation of Shanghai (grant 16ZR1442200), and the Key
Research Program of Frontier Sciences of CAS (No. QYZDJSSW-
SYS008).  This work made use of the High Performance Computing Resource in the Core
Facility for Advanced Research Computing at Shanghai Astronomical
Observatory.

\end{document}